\documentclass[aps,prb,showpacs,floatfix,twocolumn,byrevtex,superscriptaddress]{revtex4-1}
\usepackage{amsmath}
\usepackage{amssymb}
\usepackage{amstext}
\usepackage{amsopn}
\usepackage{amsfonts}
\usepackage{amsxtra}
\usepackage[english]{babel}
\usepackage{amsmath}
\usepackage{graphicx}
\usepackage{float}
\usepackage{bm}
\usepackage{multirow}
\usepackage{dcolumn}
\usepackage{hyperref}

\newcommand{\icm}{\ensuremath{~\textrm{cm}^{-1}}}% % cm-1

\newcommand{\CNFA}{Ca$_{0.77}$Nd$_{0.23}$FeAs$_{2}$}
\newcommand{\CLFA}{Ca$_{0.73}$La$_{0.27}$FeAs$_{2}$}
\newcommand{\CLFAx}{Ca$_{1-x}$La$_{x}$FeAs$_{2}$}
\newcommand{\CRFAx}{Ca$_{1-x}$RE$_{x}$FeAs$_{2}$}
\newcommand{\CNFAx}{Ca$_{1-x}$Nd$_{x}$FeAs$_{2}$}
\newcommand{\BFA}{BaFe$_{2}$As$_{2}$}

\begin{document}
\title{Optical study of the antiferromagnetic ordered state in electron-overdoped \CNFA}
\author{Run Yang}
\affiliation{Beijing National Laboratory for Condensed Matter Physics, Institute of Physics, Chinese Academy of Sciences, P.O. Box 603, Beijing 100190, China}
\author{Bing Xu}
\affiliation{Center for High Pressure Science and Technology Advanced Research, Beijing 100094, China}
\author{Yaomin Dai}
\affiliation{Center for Integrated Nanotechnologies, Los Alamos National Laboratory, Los Alamos, New Mexico 87545, USA}

\author{Wei Zhang}
\affiliation{Beijing National Laboratory for Condensed Matter Physics, Institute of Physics, Chinese Academy of Sciences, P.O. Box 603, Beijing 100190, China}
\affiliation{College of Physics, Optoelectronics and Energy \&\ Collaborative Innovation Center of Suzhou Nano Science and Technology, Soochow University, Suzhou 215006, China}
\author{Jinyun Liu}
\author{Ziyang Qiu}
\affiliation{Beijing National Laboratory for Condensed Matter Physics, Institute of Physics, Chinese Academy of Sciences, P.O. Box 603, Beijing 100190, China}
\author{Xianggang Qiu}
\email[]{xgqiu@iphy.ac.cn}
\affiliation{Beijing National Laboratory for Condensed Matter Physics, Institute of Physics, Chinese Academy of Sciences, P.O. Box 603, Beijing 100190, China}
\affiliation{Collaborative Innovation Center of Quantum Matter, Beijing 100084, China}
\date{\today}
%%%%%%%%%%%%%%%%%%%%%%%%%%%%%%%%%%%%
%
% Abstract
%

\begin{abstract}
In \CRFAx\ (RE= rare earth), an antiferromagnetic (AFM) phase as well as a structural transition has been reported, even in the electron-overdoped regime. Here we investigated the temperature-dependent in-plane optical spectroscopy of overdoped \CNFA. Upon entering the AFM state, we found an abrupt reduction of low-frequency (500-2\,000\icm) spectral weight in the optical conductivity. In sharp contrast to the parent compounds of 122 system, where spin-density-wave gaps have been clearly observed in the AFM state, a gap signature is absent in \CNFA. This may be a consequence of the poor nesting condition between hole and electron pockets. However, a spectral weight analysis shows that the reduced spectral weight at low frequency is transferred to the high frequency range ($\gtrsim 4\,000$\icm), pointing to a localization effect. These observations suggest that the AFM order in \CNFA\ is most likely to originate from a localized nature rather than Fermi surface nesting.
\end{abstract}

%  72.15.-v  Electronic conduction in metals and alloys
%  74.70.-b  SC: Superconducting materials other than cuprates
%  78.20.-e  Optical properties of bulk materials and thin films
%  78.30.-j  Infrared and Raman spectra

\maketitle

%%%%%%%%%%%%%%%%%%%%%%%%%%%%%%%%%%%%%%%%%%%%%%%%%%%%%%%%%%%%%%%%%%%%%%%%%%%%%%%
%
% Introduction

Since unconventional superconductivity usually arises as a consequence of suppressing an AFM order in the parent compounds by chemical substitution, its pairing mechanism is believed to be intimately related to the magnetism~\cite{Lee2006}. Understanding the origin of magnetism may provide important clue to the unconventional pairing. The parent compounds of iron-based superconductors (FeSCs), such as \BFA~\cite{Richard2010} and LaFeAsO~\cite{Yildirim2008}, generally feature well nested hole/electron pockets and undergoes AFM transition at low temperature~\cite{Chen2014}. Charge doping can degrade the nesting condition to suppress the AFM transition, thus inducing superconductivity at the boundary of AFM phase. Therefore, the paring mechanism has been proposed to be associate with the scattering between hole and electron pockets, which is induced by the nesting related spin-density-wave- (SDW-) type fluctuations~\cite{Chen2014,Mazin2008}. In this picture, magnetism, as well as superconductivity, may strongly depend on the topology of the Fermi surface~\cite{Yin2011}.

However, in FeTe the Fermi surface nesting fails in accounting for the bicollinear AFM order~\cite{Li2009}. Moreover, in K$_{2}$Fe$_{4}$Se$_{5}$~\cite{Mou2011} and (Li$_{0.8}$Fe$_{0.2}$)(OH)FeSe~\cite{Zhao2016}, even though no hole pockets exist at $\Gamma$ point, the superconducting gaps still open on the Fermi surfaces. Thus, the pairing mechanism cannot be understood in terms of Fermi surface instability, the electrons below the Fermi surface may also play an important role~\cite{Chen2014}. Moreover, Iimura et al.~\cite{Hiraishi2014,Iimura2012} has found another superconducting dome and AFM ordered state in electron-overdoped area of REFeAsO$_{1-x}$H$_{x}$ (RE= rare earth), these anomalous behaviors introduce additional puzzle and could reveal more rich physics in FeSCs. However, since the lack of large sized single crystal, there is difficulty in carrying out detail experiments.

\CRFAx\ is a newly discovered iron-pnictide family with a monotonic structure (space group $P_{21}$)~\cite{Katayama2013}. It consists of alternatively stacked FeAs and (Ca,RE)As layers. Rare earth doping on Ca site can introduce electrons into the FeAs layers and induce superconductivity with $T_{c}=35~K (x=0.15)$~\cite{Li2015}. With further electron doping, superconductivity is suppressed, however instead of the Fermi-liquid (FL) behavior in some other iron pnictides, there is a recovery of the AFM order, which is confirmed by recent nuclear magnetic resonance (NMR)~\cite{Kawasaki2015} and neutron scattering results~\cite{Jiang2016}, similar to what happened in REFeAsO$_{1-x}$H$_{x}$~\cite{Hiraishi2014}. Since the AFM is intimately related to superconductivity, studying such anomalous behavior may provide some new physics and shed new light on the mechanism of the unconventional superconductivity in FeSCs .

In this work, we have synthesized single crystals of the electron overdoped \CNFA~\cite{Katayama2013} ($T_{N}\lesssim$73~K), and measured its temperature dependent in-plane reflectivity before and after the AFM phase transition. To analysis the optical conductivity, we use two Drude components to describe the low energy ($<2\,000$\icm) response, to account for the multiband nature of FeSCs~\cite{Wu2010,Dai2015}. Such two-Drude model reveals a narrow temperature-dependent Drude item representing the coherent response, and a broad temperature-independent Drude one representing the incoherent response. Across the phase transition, we notice that the broad Drude component is significantly suppressed. Correspondingly, the spectral weight of optical conductivity at low frequency (50-2\,000\icm) is reduced. Unlike the parent compound of 122 system, a lack of the signature of SDW transition in the optical conductivity indicates the poor nesting condition. However, a spectral weight analysis shows that the reduced spectral weight at low frequency is transferred to the high frequency area ($\gtrsim 4\,000$\icm), indicating a localization effect, such as Hund's rule coupling. We infer that the magnetism in such electron-overdoped system may come from a localization effect and do not rely on the topology of the Fermi surface.

%%%%%%%%%%%%%%%%%%%%%%%%%%%%%%%%%%%%%%%%%%%%%%%%%%%%%%%%%%%%%%%%%%%%%%%%%%%%%%%
%
% Experiments
%
High-quality single crystal of \CNFAx\ were synthesized by heating a mixture of Ca, Nd, FeAs, As powders with nominal composition of $x=0.2$~\cite{Katayama2013,Kudo2014}, typical size was about 3$\times$3$\times$0.1~mm$^3$. The composition determined by Inductive Coupled Plasma Emission Spectrometer (ICP) was \CNFA. Resistivity measurement was taken on Quantum Design Physical Property Measurement System (PPMS). Magnetization was measured using a Quantum Design superconducting quantum interference device (SQUID).
The reflectivity from the cleaved surface has been measured at a near-normal angle of incidence on a Fourier transform infrared spectrometer (Bruker 80v) for light polarized in the $ab$ planes using an $in~situ$ evaporation technique.~\cite{Homes1993} Data from 40 to 15\,000\icm\ were collected at 8 different temperatures from 15 to 300~K on an ARS-Helitran cryostat. The reflectivity in visible and UV range (10\,000-40\,000\icm) at room temperature was taken with an Avaspec 2048$\times$14 optical fiber spectrometer. The optical conductivity has been determined from a Kramers-Kronig analysis of reflectivity $R(\omega)$ over the entire frequency range. Since the measurement is made in a limited energy range. A Hagen-Rubens relation ($R=1-A\sqrt{\omega}$) is used for low-frequency extrapolation. Above the highest-measured frequency (40\,000\icm), $R(\omega)$ is assumed to be constant up to 40~eV, above which a free-electron response ($\omega^{-4}$) is used~\cite{Dai2013}.

%
% Figure 1
%

\begin{figure}[tb]
\includegraphics[width=0.9\columnwidth]{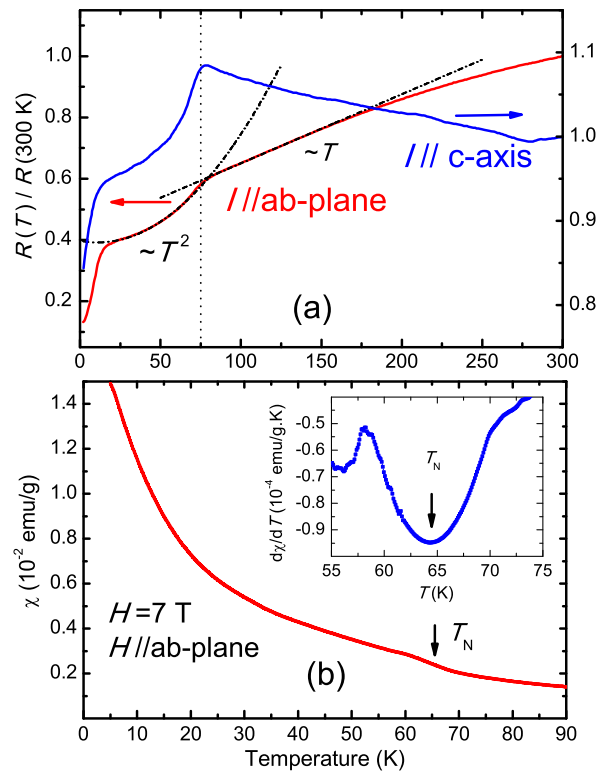}
\caption{ (color online) (a) Temperature dependence of normalized resistivity of \CNFA with $I\parallel c-axis$ (blue) and $I\parallel ab-plane$ (red). (b) The temperature dependence of magnetic suseptbility of \CNFA with $H\parallel ab$ plane at $1$~T. Inset is its derivative with temperature. Arrows indicate the Neel temperature.}
\label{RT}
\end{figure}

%
% Figure 2
%
\begin{figure}[tb]
\includegraphics[width=0.8\columnwidth]{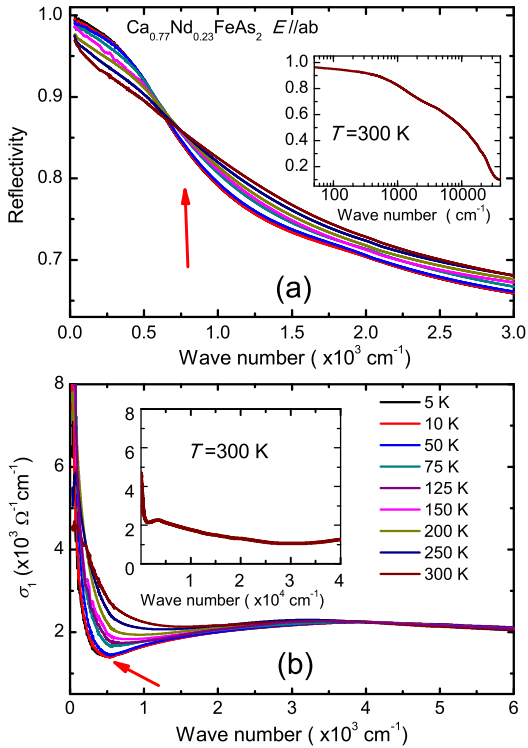}
\caption{ (color online)Reflectivity (a) and optical conductivity (b) of \CNFA\  at various temperatures. The red arrows indicates the suppression in reflectivity and optical conductivity. }
\label{sigma1}
\end{figure}
%%%%%%%%%%%%%%%%%%%%%%%%%%%%%%%%%%%%%%%%%%%%%%%%%%%%%%%%%%%%%%%%%%%%%%%%%%%%%%%
%
% Results
%
Fig.~\ref{RT}(a) shows the temperature dependent resistance with $I\parallel ab$ plane and $I\parallel c$ axis, we note that above 73~K the in-plane resistance shows a metallic behavior which decreases upon cooling, while the inter-plane resistance shows an insulating behavior, i.e., increases with decreasing temperature. It means that even though the CaAs layer has been reported to be metallic, the inter-layer coupling is still weak. While at 73~K, we note a sudden drop on both curves, and a kink could also be observed in its magnetic susceptibility [Fig.~\ref{RT}(b)], these anomalies have been demonstrated to be caused by structural and AFM phase transition~\cite{Jiang2016}. Below 10~K, another drop in $R(T)$ [Fig.~\ref{RT}(a)] may come from the filamentary superconductivity. In addition to these observations, we also note that, in the normal state, the in-plane resistance shows  an obvious non-Fermi-liquid (NFL) behavior(discrepancy of the $T^{2}$ behavior)~\cite{Analytis2014,Dai2015}, but after the phase transition the system becomes much more coherent. Such phenomenon can be understood in terms of the spin-fluctuations scattering, i.e., in the normal state, there exists strong spin fluctuation to scatter the carrier and results in NFL behavior, while in the magnetic ordered state the fluctuation is greatly suppressed~\cite{Dai2015}.

The measured in-plane reflectivity and the real part of the optical conductivity $\sigma_1(\omega,~T)$ are summarized in Fig.~\ref{sigma1}, for selected temperature above and below the AFM phase transition. The reflectivity shows a typical metallic behavior, approaching unity at low frequencies ($<1\,000\icm$) and increasing upon cooling. After the AFM phase transition, a suppression of $R(\omega,~T)$ in the mid-infrared range (500-2\,000\icm) can be seen. Correspondingly, similar behavior is also found in the optical conductivity spectra, indicating the change of the band structure and a loss of density of states near the Fermi level. Comparing with \BFA, which also undergoes an AFM phase transition~\cite{Hu2008}, below 73~K, no additional gaplike feature below 3\,000\icm\ can be found, indicating no SDW transition happened in \CNFA, the reason will be discussed later.

\begin{figure}[tb]
\centering
\includegraphics[width=0.8\columnwidth]{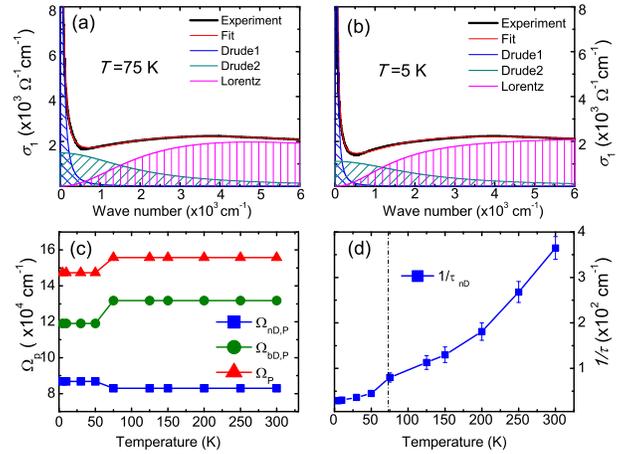}
\caption{(color on line) (a) (b) Optical conductivity of \CNFA\ at 75~K  and 5~K (thick black lines), fitting with the Drude-Lorentz model (thin red lines) and its decomposition into individual Drude and Lorentz terms. (c) Plasma frequency  for the narrow Drude(blue), broad Drude(green) and the total weights of the narrow and broad Drude components (red) for various temperatures. (d) $T$ dependence of the quasiparticle scattering rate $1/\tau_{nD}$ derived from the coherent narrow Drude component.(The vertical dashed line denotes the phase transition.) }
\label{fit}
\end{figure}

To quantitatively analyze the optical data of \CNFA, we fit the $\sigma_1(\omega,T)$ with a simple Drude-Lorentz mode for the dielectric function:
\begin{equation}
\label{DrudeLorentz}
\epsilon(\omega)=\epsilon_{\infty}-\sum_{i}\frac{\Omega^{2}_{p,i}}{\omega^{2}+\frac{i\omega}{\tau_{i}}}+\sum_{j}\frac{\Omega^{2}_{j}}{\omega^2_{j}-\omega^2-\frac{i\omega}{\tau_{i}}},
\end{equation}
where $\epsilon_{\infty}$ is the real part of the dielectric function at high frequencies, the second term corresponds to the Drude response characterized by a plasma frequency $\Omega_{p,i}^{2}=4\pi ne^{2}/m^{*}$, with $n$ a carrier concentration and $m^{*}$ an effective mass, and $1/\tau_i$ the scattering rate. The third term is a sum of Lorentz oscillators characterized by a resonance frequency $\omega_{j}$, a linewidth $\gamma_{j}$, and an oscillator strength $\Omega_{j}$. The Drude term accounts for the itinerant carrier (intraband) response, while the Lorentz contributions represent the localized (interband) excitations~\cite{Homes2015}. The complex conductivity $\tilde{\sigma}(\omega)=\sigma_{1}+i\sigma_{2}=i\omega[\tilde{\varepsilon}(\omega)-\varepsilon_{\infty}]/60$ (in units of $\Omega^{-1}\cdot\icm$).  The fitting results shown in Fig.~\ref{fit}(a)(b) can well reproduce the experimental results.

Fig.~\ref{fit}(a)(b) summarize the fitting results, considering the multiband nature of FeSCs, we use two Drude component(a narrow one and a broad one) and a Lorentz components to describe the optical response below 6\,000\icm~\cite{Wu2010}. The narrow Drude item has strongly temperature-dependent scattering rate at low temperature($1/\tau_{nD}\simeq32\icm$, 10~K) and represents the coherent response. Whereas the scattering rate of the broad Drude item is 2\,000\icm or larger, corresponding to a mean free path shorter than the lattice spacing and indicating a highly incoherent character~\cite{Nakajima2014,Nakajima2014a}. The broad Drude item is much stronger than the narrow one, but it almost does not change with the temperature, representing an incoherent background~\cite{Homes2015}. Across the AFM transition, in Fig.~\ref{fit}(c) we note the broad Drude component is greatly suppressed. Meanwhile, the scattering rate of the coherent response is also reduced [Fig.~\ref{fit}(d)]. Considering the suppressed spin fluctuation in AFM ordered state, we infer the incoherent response may have a relation with the AFM fluctuation. The overall plasma frequency $\Omega_{p}$ is considered to contribute from both narrow and broad Drude component with $\Omega_{p}=(\Omega_{nD,p}^{2}+\Omega_{bD,p}^{2})^{1/2}$. In Fig.~\ref{fit}(c), we note that even though the narrow Drude component is a little enhanced, the overall frequency is suppressed across the phase transition, indicating a loss of the itinerant carriers after the transition. So where do the lost carriers go?

% Figure 3
%

\begin{figure}[tb]
\centering
\includegraphics[width=0.75\columnwidth]{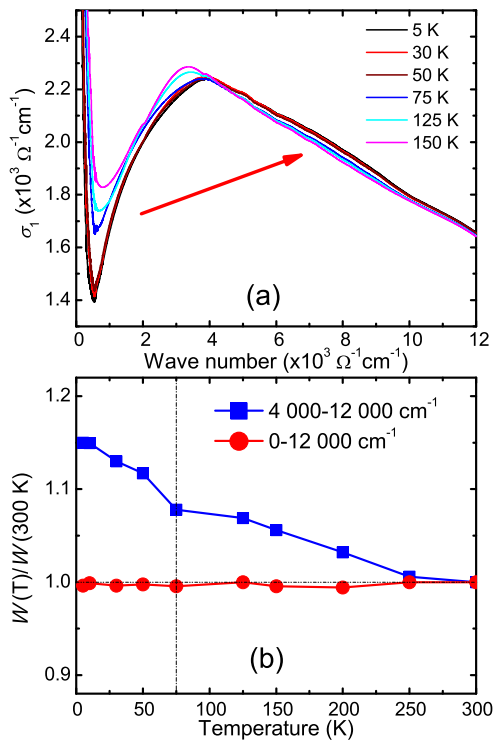}
\caption{(color on line) (a) Enlarged optical conductivity of \CNFA\ before and after the phase transition, from which we could clearly see the spectral weight transference indicated by the red arrow. (b) Temperature dependence of the spectral weight, $W^{\omega_b}_{\omega_a}~=~\int^{\omega_b}_{\omega_a}\sigma_1(\omega)d\omega$ between different lower and upper cutoff frequencies. The vertical dashed line denotes the phase transition.}
\label{SW}
\end{figure}

In Fig.~\ref{SW}(b), we have calculated the spectral weight in different areas of optical conductivity. From the results, one notes that the spectral weight at high energy (4\,000-15\,000\icm) varies significantly with the temperature. Upon cooling, the spectral weight at high frequency is continuously enhanced while the overall spectral weight keeps constant. It means the spectral weight at low frequency is transferred from low to high frequency area. Such a pseudogap-like behavior has been widely attributed to the Hund's rule coupling effect, which can localize and polarize the itinerant electrons to enhance the electron correlation and AFM exchange interaction~\cite{Wang2012}. At low temperature ($\lesssim73~K$), we observed an additional enhancement of the high-frequency spectral weight, combined with the result of fitting analysis, we find that the suppressed spectral weight at low frequency area (500-2\,000\icm) in the magnetic ordered state has been moved to the high frequency area, reflecting a sudden enhanced localization effect. Further research is needed to determine whether this effect is caused by Hund's rule coupling or some other effects.

% Figure 4
%

%%%%%%%%%%%%%%%%%%%%%%%%%%%%%%%%%%%%%%%%%%%%%%%%%%%%%%%%%%%%%%%%%%%%%%%%%%%%%%%
%
% Discussion

Even though the spectral weight at low frequency area is greatly transferred to high frequency area during the phase transition, we do not observe any additional gap-like feature in the optical conductivity (Fig. ~\ref{sigma1}(b)) after the AFM phase transition (evidence given by spectral analysis is shown in Appendix B). This is different from what happened in \BFA, which undergoes an SDW transition and has a signature of opening a gap in its optical spectrum~\cite{Hu2008}. Comparing their Fermi surface, we note that, \BFA\ as a compensate metal has equal sized electron and hole pockets, and there is a good nesting between them along the ($\pi,\pi$) direction in the Brillouin zone~\cite{Richard2011,Richard2010}. This will lead to the instability of the Fermi surface and SDW transition at low temperature~\cite{Hancock2010}. Whereas recent ARPES results on \CLFA\ indicate electron-overdoped FeAs layers~\cite{Jiang2016}, in which electron pockets is much bigger than hole pockets and lead to poor nesting condition (comparing with the less doped ones) so as to no SDW transition and no gap on its Fermi surface. The AFM transition in such poorly nested system suggests the magnetism in electron-overdoped \CNFA\ does not rely on the topology of the Fermi surface. The spectral weight transfer without a SDW gap creates a pseudogap feature in the optical spectrum across the AFM phase transition.

Such strong magnetism in electron-overdoped \CRFAx\ cannot be understood by previous theoretical model~\cite{Dai2012,Ikeda2010}, in which the electron doping could continuously suppress the magnetic order and low energy spin fluctuation with worsening nesting condition. Moreover, recent NMR results point out that magnetism is rather enhanced with doping~\cite{Kawasaki2015}, such behavior cannot be understood in terms of the Fermi surface nesting. Thus, the mechanism for the magnetism in overdoped area should be reconsidered.

Very recently, the XRD experiments~\cite{Jiang2016}  on \CLFAx\ found that the height of As atom (in FeAs layer) corresponding to Fe layer in electron-overdoped sample ($x=0.27$) is 1.422(5)~{\AA}, while it is 1.412(5)~{\AA} for the less doped one ($x=0.195$)~\cite{Katayama2013}, indicating the rare earth doping could somehow increase the pnictogen height in \CLFAx. Such behavior could enlarge the distance between Fe and As atoms and suppress the hybridization between Fe 3d and As 4p orbitals~\cite{diehl}. As a result, electrons on Fe atoms could be much more localized and the electron correlation as well as the local magnetic moment are enhanced~\cite{Zhang2014,diehl}. Furthermore, across the AFM phase transition in Ca$_{0.73}$La$_{0.27}$FeAs$_{2}$, the lattice parameters a,b,c as well as the cell volume are found to be enlarged abruptly, such negative thermal expansion could narrow the bandwidth and localize the electrons, which is consistent with our observations in optical spectroscopy, thus we propose that the magnetic transition is mostly like to have a localized origin. Even though an enhanced localization effect accompany with the phase transition, the system is still in a metallic state. From the fitting analysis of the optical data in Fig.~\ref{fit}(a)(b), we note that the AFM transition mainly affects the broad Drude item, while the narrow one is almost unaffected. This may reflect the multiband nature of FeSCs~\cite{Si2009,DeMedici2009}, from which the much incoherent bands (or orbitals) are selected to be localized across the phase transition~\cite{Neupane2009}.

%%%%%%%%%%%%%%%%%%%%%%%%%%%%%%%%%%%%%%%%%%%%%%%%%%%%%%%%%%%%%%%%%%%%%%%%%%%%%%%
%
% Conclusions
%

In summary, we have synthesized the single crystal of \CNFA\ and investigated its optical response at temperatures before and after the phase transition. From the optical conductivity, we find obvious spectral weight transfer from low to high energy area without any signature of SDW gap across the AFM phase transition. Therefore, we propose the magnetism in electron-overdoped \CRFAx\ may come from the localization effect, which results in a pseudogap feature, and do not rely on the Fermi surface nesting. Since further electron doping by Co in Ca$_{0.73}$La$_{0.27}$Fe$_{1-x}$Co$_{x}$As$_{2}$ could induce another superconducting dome~\cite{Jiang2015}, more investigation on this system may shed new lights on the mechanism of high-$T_{c}$ superconductivity in FeSCs.

%%%%%%%%%%%%%%%%%%%%%%%%%%%%%%%%%%%%%%%%%%%%%%%%%%%%%%%%%%%%%%%%%%%%%%%%%%%%%%%
%
% Acknowledgment
%

\begin{acknowledgments}
We thank Kui Jin, Shiliang Li, Congcong Le and Xianxin Wu for useful discussion.
This work was supported by the NSFC (Grants No. 91121004 and 973 Projects No. 2015CB921303) and the MSTC (973 Projects No. 2011CBA00107, No. 2012CB821400, No. 2012CB921302, and No. 2009CB929102).
\end{acknowledgments}

%%%%%%%%%%%%%%%%%%%%%%%%%%%%%%%%%%%%%%%%%%%%%%%%%%%%%%%%%%%%%%%%%%%%%%%%%%%%%%%
%
% The Appendix (BibTeX)
%
\appendix

\section{Hall coefficient}

To determine the electron concentration of \CNFA\ we measured its Hall resistance $\rho_{xy}$ and calculated the Hall coefficient at 300~K. The Hall coefficient of \CNFA at 300~K is $-9.5\times10^{-9} m^{3}/C$, which is comparable with that of \CLFA\ ~\cite{Jiang2016a}, indicating that the \CNFA\ is electron-overdoped.

\begin{figure}[H]
\includegraphics[width=0.6\columnwidth]{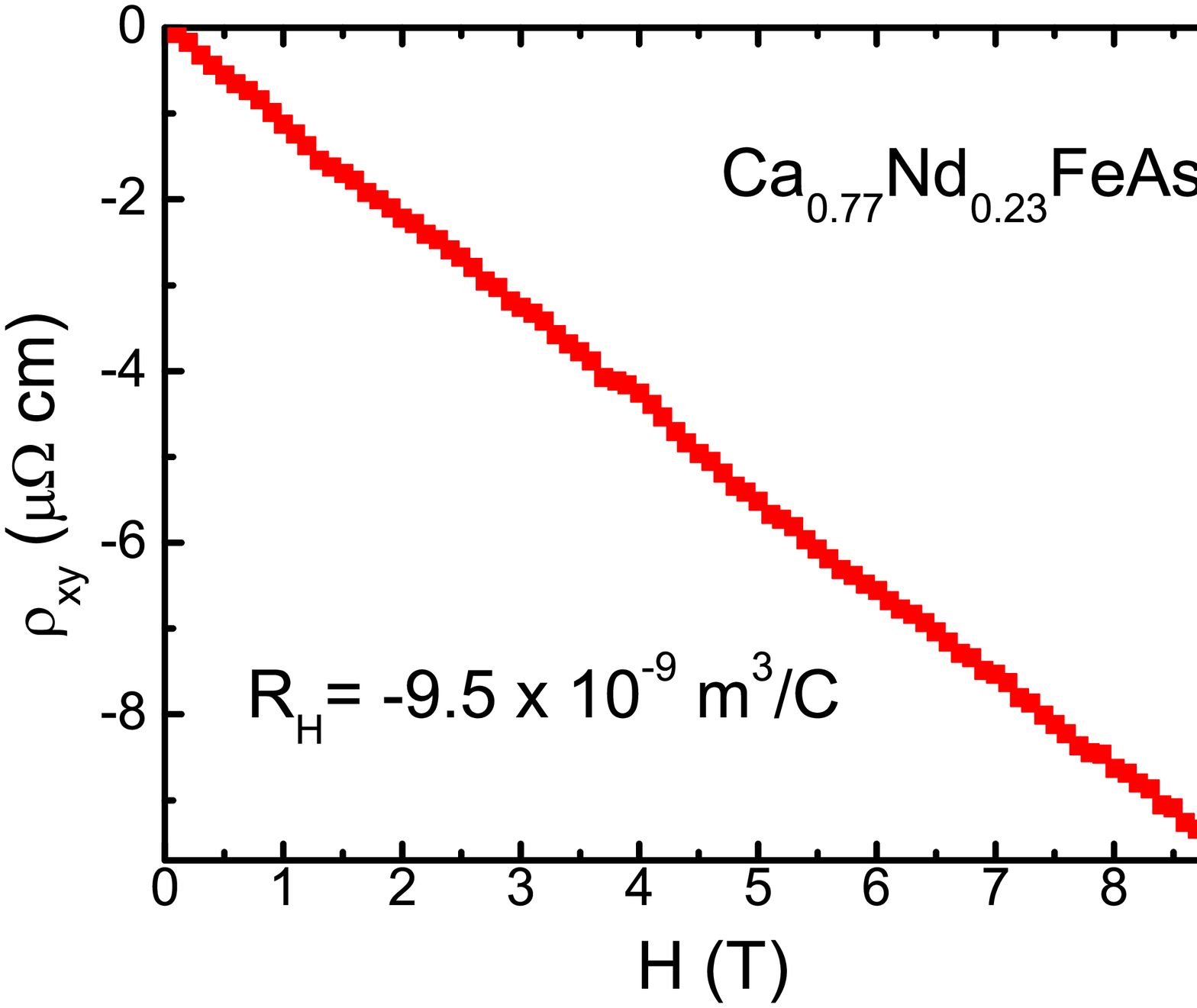}
\caption{ (color online)Hall resistance and Hall coefficient of \CNFA\  at 300 K.  }
\label{Hall}
\end{figure}

\section{Detail spectral weight analysis}

\begin{figure}[H]
\includegraphics[width=0.7\columnwidth]{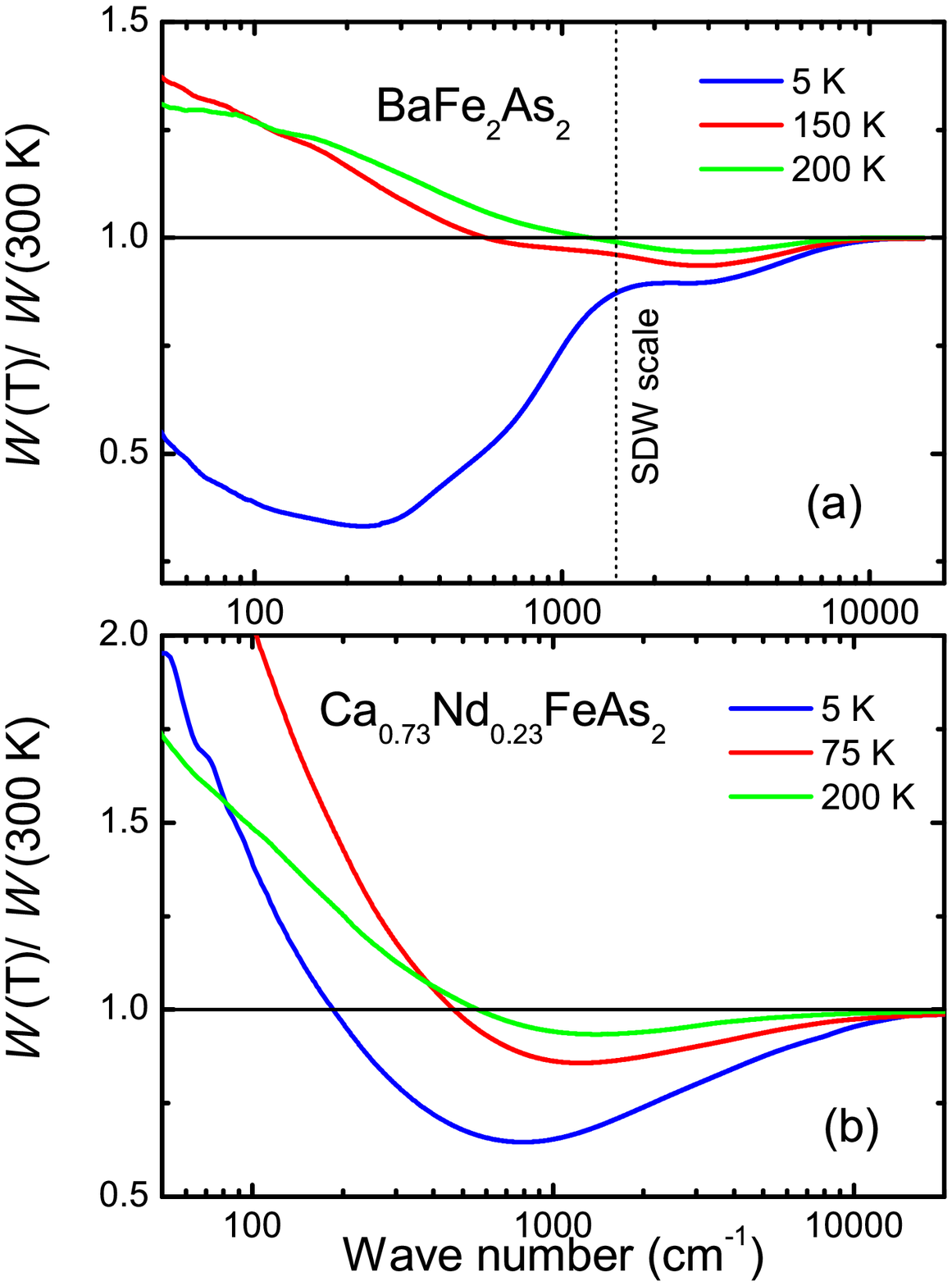}
\caption{ (color online)Ratio of the integrated spectral weight W(T)/W(300) as a function of cutoff frequency (wave number) in \BFA\ (a) and \CNFA\ (b). }
\label{SW1}
\end{figure}

In Fig.~\ref{SW1} we calculated the ratio of the spectral weight as a function of cutoff frequency for \BFA\ and \CNFA\ respectively. Comparing with \BFA, we find no obvious signature characteristic of opening a gap within the SDW scale of \BFA\ but an enhanced spectral weight transfer from low to high energy area (Fig.~\ref{SW}(a)(b)), such behavior might be regarded as a pseudogap behavior.

%%%%%%%%%%%%%%%%%%%%%%%%%%%%%%%%%%%%%%%%%%%%%%%%%%%%%%%%%%%%%%%%%%%%%%%%%%%%%%%
%
%The bibliography (BibTeX)
%
%\bibliography{bib}

\begin{thebibliography}{37}%
\makeatletter
\providecommand \@ifxundefined [1]{%
 \@ifx{#1\undefined}
}%
\providecommand \@ifnum [1]{%
 \ifnum #1\expandafter \@firstoftwo
 \else \expandafter \@secondoftwo
 \fi
}%
\providecommand \@ifx [1]{%
 \ifx #1\expandafter \@firstoftwo
 \else \expandafter \@secondoftwo
 \fi
}%
\providecommand \natexlab [1]{#1}%
\providecommand \enquote  [1]{``#1''}%
\providecommand \bibnamefont  [1]{#1}%
\providecommand \bibfnamefont [1]{#1}%
\providecommand \citenamefont [1]{#1}%
\providecommand \href@noop [0]{\@secondoftwo}%
\providecommand \href [0]{\begingroup \@sanitize@url \@href}%
\providecommand \@href[1]{\@@startlink{#1}\@@href}%
\providecommand \@@href[1]{\endgroup#1\@@endlink}%
\providecommand \@sanitize@url [0]{\catcode `\\12\catcode `\$12\catcode
  `\&12\catcode `\#12\catcode `\^12\catcode `\_12\catcode `\%12\relax}%
\providecommand \@@startlink[1]{}%
\providecommand \@@endlink[0]{}%
\providecommand \url  [0]{\begingroup\@sanitize@url \@url }%
\providecommand \@url [1]{\endgroup\@href {#1}{\urlprefix }}%
\providecommand \urlprefix  [0]{URL }%
\providecommand \Eprint [0]{\href }%
\providecommand \doibase [0]{http://dx.doi.org/}%
\providecommand \selectlanguage [0]{\@gobble}%
\providecommand \bibinfo  [0]{\@secondoftwo}%
\providecommand \bibfield  [0]{\@secondoftwo}%
\providecommand \translation [1]{[#1]}%
\providecommand \BibitemOpen [0]{}%
\providecommand \bibitemStop [0]{}%
\providecommand \bibitemNoStop [0]{.\EOS\space}%
\providecommand \EOS [0]{\spacefactor3000\relax}%
\providecommand \BibitemShut  [1]{\csname bibitem#1\endcsname}%
\let\auto@bib@innerbib\@empty
%</preamble>
\bibitem [{\citenamefont {Lee}\ \emph {et~al.}(2006)\citenamefont {Lee},
  \citenamefont {Nagaosa},\ and\ \citenamefont {Wen}}]{Lee2006}%
  \BibitemOpen
  \bibfield  {author} {\bibinfo {author} {\bibfnamefont {P.~A.}\ \bibnamefont
  {Lee}}, \bibinfo {author} {\bibfnamefont {N.}~\bibnamefont {Nagaosa}}, \ and\
  \bibinfo {author} {\bibfnamefont {X.-G.}\ \bibnamefont {Wen}},\ }\href
  {\doibase 10.1103/RevModPhys.78.17} {\bibfield  {journal} {\bibinfo
  {journal} {Reviews of Modern Physics}\ }\textbf {\bibinfo {volume} {78}},\
  \bibinfo {pages} {17} (\bibinfo {year} {2006})}\BibitemShut {NoStop}%
\bibitem [{\citenamefont {Richard}\ \emph {et~al.}(2010)\citenamefont
  {Richard}, \citenamefont {Nakayama}, \citenamefont {Sato}, \citenamefont
  {Neupane}, \citenamefont {Xu}, \citenamefont {Bowen}, \citenamefont {Chen},
  \citenamefont {Luo}, \citenamefont {Wang}, \citenamefont {Dai}, \citenamefont
  {Fang}, \citenamefont {Ding},\ and\ \citenamefont {Takahashi}}]{Richard2010}%
  \BibitemOpen
  \bibfield  {author} {\bibinfo {author} {\bibfnamefont {P.}~\bibnamefont
  {Richard}}, \bibinfo {author} {\bibfnamefont {K.}~\bibnamefont {Nakayama}},
  \bibinfo {author} {\bibfnamefont {T.}~\bibnamefont {Sato}}, \bibinfo {author}
  {\bibfnamefont {M.}~\bibnamefont {Neupane}}, \bibinfo {author} {\bibfnamefont
  {Y.-M.}\ \bibnamefont {Xu}}, \bibinfo {author} {\bibfnamefont {J.~H.}\
  \bibnamefont {Bowen}}, \bibinfo {author} {\bibfnamefont {G.~F.}\ \bibnamefont
  {Chen}}, \bibinfo {author} {\bibfnamefont {J.~L.}\ \bibnamefont {Luo}},
  \bibinfo {author} {\bibfnamefont {N.~L.}\ \bibnamefont {Wang}}, \bibinfo
  {author} {\bibfnamefont {X.}~\bibnamefont {Dai}}, \bibinfo {author}
  {\bibfnamefont {Z.}~\bibnamefont {Fang}}, \bibinfo {author} {\bibfnamefont
  {H.}~\bibnamefont {Ding}}, \ and\ \bibinfo {author} {\bibfnamefont
  {T.}~\bibnamefont {Takahashi}},\ }\href {\doibase
  10.1103/PhysRevLett.104.137001} {\bibfield  {journal} {\bibinfo  {journal}
  {Phys. Rev. Lett.}\ }\textbf {\bibinfo {volume} {104}},\ \bibinfo {pages}
  {137001} (\bibinfo {year} {2010})}\BibitemShut {NoStop}%
\bibitem [{\citenamefont {Yildirim}(2008)}]{Yildirim2008}%
  \BibitemOpen
  \bibfield  {author} {\bibinfo {author} {\bibfnamefont {T.}~\bibnamefont
  {Yildirim}},\ }\href {\doibase 10.1103/PhysRevLett.101.057010} {\bibfield
  {journal} {\bibinfo  {journal} {Phys. Rev. Lett.}\ }\textbf {\bibinfo
  {volume} {101}},\ \bibinfo {pages} {057010} (\bibinfo {year}
  {2008})}\BibitemShut {NoStop}%
\bibitem [{\citenamefont {Chen}\ \emph {et~al.}(2014)\citenamefont {Chen},
  \citenamefont {Dai}, \citenamefont {Feng}, \citenamefont {Xiang},\ and\
  \citenamefont {Zhang}}]{Chen2014}%
  \BibitemOpen
  \bibfield  {author} {\bibinfo {author} {\bibfnamefont {X.}~\bibnamefont
  {Chen}}, \bibinfo {author} {\bibfnamefont {P.}~\bibnamefont {Dai}}, \bibinfo
  {author} {\bibfnamefont {D.}~\bibnamefont {Feng}}, \bibinfo {author}
  {\bibfnamefont {T.}~\bibnamefont {Xiang}}, \ and\ \bibinfo {author}
  {\bibfnamefont {F.-C.}\ \bibnamefont {Zhang}},\ }\href {\doibase
  10.1093/nsr/nwu007} {\bibfield  {journal} {\bibinfo  {journal} {National
  Science Review}\ }\textbf {\bibinfo {volume} {1}},\ \bibinfo {pages} {371}
  (\bibinfo {year} {2014})}\BibitemShut {NoStop}%
\bibitem [{\citenamefont {Mazin}\ \emph {et~al.}(2008)\citenamefont {Mazin},
  \citenamefont {Singh}, \citenamefont {Johannes},\ and\ \citenamefont
  {Du}}]{Mazin2008}%
  \BibitemOpen
  \bibfield  {author} {\bibinfo {author} {\bibfnamefont {I.~I.}\ \bibnamefont
  {Mazin}}, \bibinfo {author} {\bibfnamefont {D.~J.}\ \bibnamefont {Singh}},
  \bibinfo {author} {\bibfnamefont {M.~D.}\ \bibnamefont {Johannes}}, \ and\
  \bibinfo {author} {\bibfnamefont {M.~H.}\ \bibnamefont {Du}},\ }\href
  {\doibase 10.1103/PhysRevLett.101.057003} {\bibfield  {journal} {\bibinfo
  {journal} {Phys. Rev. Lett.}\ }\textbf {\bibinfo {volume} {101}},\ \bibinfo
  {pages} {057003} (\bibinfo {year} {2008})}\BibitemShut {NoStop}%
\bibitem [{\citenamefont {Yin}\ \emph {et~al.}(2011)\citenamefont {Yin},
  \citenamefont {Haule},\ and\ \citenamefont {Kotliar}}]{Yin2011}%
  \BibitemOpen
  \bibfield  {author} {\bibinfo {author} {\bibfnamefont {Z.~P.}\ \bibnamefont
  {Yin}}, \bibinfo {author} {\bibfnamefont {K.}~\bibnamefont {Haule}}, \ and\
  \bibinfo {author} {\bibfnamefont {G.}~\bibnamefont {Kotliar}},\ }\href
  {\doibase 10.1038/nmat3120} {\bibfield  {journal} {\bibinfo  {journal}
  {Nature Materials}\ }\textbf {\bibinfo {volume} {10}},\ \bibinfo {pages}
  {932} (\bibinfo {year} {2011})}\BibitemShut {NoStop}%
\bibitem [{\citenamefont {Li}\ \emph {et~al.}(2009)\citenamefont {Li},
  \citenamefont {de~la Cruz}, \citenamefont {Huang}, \citenamefont {Chen},
  \citenamefont {Lynn}, \citenamefont {Hu}, \citenamefont {Huang},
  \citenamefont {Hsu}, \citenamefont {Yeh}, \citenamefont {Wu},\ and\
  \citenamefont {Dai}}]{Li2009}%
  \BibitemOpen
  \bibfield  {author} {\bibinfo {author} {\bibfnamefont {S.}~\bibnamefont
  {Li}}, \bibinfo {author} {\bibfnamefont {C.}~\bibnamefont {de~la Cruz}},
  \bibinfo {author} {\bibfnamefont {Q.}~\bibnamefont {Huang}}, \bibinfo
  {author} {\bibfnamefont {Y.}~\bibnamefont {Chen}}, \bibinfo {author}
  {\bibfnamefont {J.~W.}\ \bibnamefont {Lynn}}, \bibinfo {author}
  {\bibfnamefont {J.}~\bibnamefont {Hu}}, \bibinfo {author} {\bibfnamefont
  {Y.-L.}\ \bibnamefont {Huang}}, \bibinfo {author} {\bibfnamefont {F.-C.}\
  \bibnamefont {Hsu}}, \bibinfo {author} {\bibfnamefont {K.-W.}\ \bibnamefont
  {Yeh}}, \bibinfo {author} {\bibfnamefont {M.-K.}\ \bibnamefont {Wu}}, \ and\
  \bibinfo {author} {\bibfnamefont {P.}~\bibnamefont {Dai}},\ }\href {\doibase
  10.1103/PhysRevB.79.054503} {\bibfield  {journal} {\bibinfo  {journal} {Phys.
  Rev. B}\ }\textbf {\bibinfo {volume} {79}},\ \bibinfo {pages} {054503}
  (\bibinfo {year} {2009})}\BibitemShut {NoStop}%
\bibitem [{\citenamefont {Mou}\ \emph {et~al.}(2011)\citenamefont {Mou},
  \citenamefont {Zhao},\ and\ \citenamefont {Zhou}}]{Mou2011}%
  \BibitemOpen
  \bibfield  {author} {\bibinfo {author} {\bibfnamefont {D.-x.}\ \bibnamefont
  {Mou}}, \bibinfo {author} {\bibfnamefont {L.}~\bibnamefont {Zhao}}, \ and\
  \bibinfo {author} {\bibfnamefont {X.-j.}\ \bibnamefont {Zhou}},\ }\href
  {\doibase 10.1007/s11467-011-0229-5} {\bibfield  {journal} {\bibinfo
  {journal} {Frontiers of Physics}\ }\textbf {\bibinfo {volume} {6}},\ \bibinfo
  {pages} {410} (\bibinfo {year} {2011})}\BibitemShut {NoStop}%
\bibitem [{\citenamefont {Zhao}\ \emph {et~al.}(2016)\citenamefont {Zhao},
  \citenamefont {Liang}, \citenamefont {Yuan}, \citenamefont {Hu},
  \citenamefont {Liu}, \citenamefont {Huang}, \citenamefont {He}, \citenamefont
  {Shen}, \citenamefont {Xu}, \citenamefont {Liu}, \citenamefont {Yu},
  \citenamefont {Liu}, \citenamefont {Zhou}, \citenamefont {Huang},
  \citenamefont {Dong}, \citenamefont {Zhou}, \citenamefont {Liu},
  \citenamefont {Lu}, \citenamefont {Zhao}, \citenamefont {Chen}, \citenamefont
  {Xu},\ and\ \citenamefont {Zhou}}]{Zhao2016}%
  \BibitemOpen
  \bibfield  {author} {\bibinfo {author} {\bibfnamefont {L.}~\bibnamefont
  {Zhao}}, \bibinfo {author} {\bibfnamefont {A.}~\bibnamefont {Liang}},
  \bibinfo {author} {\bibfnamefont {D.}~\bibnamefont {Yuan}}, \bibinfo {author}
  {\bibfnamefont {Y.}~\bibnamefont {Hu}}, \bibinfo {author} {\bibfnamefont
  {D.}~\bibnamefont {Liu}}, \bibinfo {author} {\bibfnamefont {J.}~\bibnamefont
  {Huang}}, \bibinfo {author} {\bibfnamefont {S.}~\bibnamefont {He}}, \bibinfo
  {author} {\bibfnamefont {B.}~\bibnamefont {Shen}}, \bibinfo {author}
  {\bibfnamefont {Y.}~\bibnamefont {Xu}}, \bibinfo {author} {\bibfnamefont
  {X.}~\bibnamefont {Liu}}, \bibinfo {author} {\bibfnamefont {L.}~\bibnamefont
  {Yu}}, \bibinfo {author} {\bibfnamefont {G.}~\bibnamefont {Liu}}, \bibinfo
  {author} {\bibfnamefont {H.}~\bibnamefont {Zhou}}, \bibinfo {author}
  {\bibfnamefont {Y.}~\bibnamefont {Huang}}, \bibinfo {author} {\bibfnamefont
  {X.}~\bibnamefont {Dong}}, \bibinfo {author} {\bibfnamefont {F.}~\bibnamefont
  {Zhou}}, \bibinfo {author} {\bibfnamefont {K.}~\bibnamefont {Liu}}, \bibinfo
  {author} {\bibfnamefont {Z.}~\bibnamefont {Lu}}, \bibinfo {author}
  {\bibfnamefont {Z.}~\bibnamefont {Zhao}}, \bibinfo {author} {\bibfnamefont
  {C.}~\bibnamefont {Chen}}, \bibinfo {author} {\bibfnamefont {Z.}~\bibnamefont
  {Xu}}, \ and\ \bibinfo {author} {\bibfnamefont {X.~J.}\ \bibnamefont
  {Zhou}},\ }\href {\doibase 10.1038/ncomms10608} {\bibfield  {journal}
  {\bibinfo  {journal} {Nature Communications}\ }\textbf {\bibinfo {volume}
  {7}},\ \bibinfo {pages} {10608} (\bibinfo {year} {2016})}\BibitemShut
  {NoStop}%
\bibitem [{\citenamefont {Hiraishi}\ \emph {et~al.}(2014)\citenamefont
  {Hiraishi}, \citenamefont {Iimura}, \citenamefont {Kojima}, \citenamefont
  {Yamaura}, \citenamefont {Hiraka}, \citenamefont {Ikeda}, \citenamefont
  {Miao}, \citenamefont {Ishikawa}, \citenamefont {Torii}, \citenamefont
  {Miyazaki}, \citenamefont {Yamauchi}, \citenamefont {Koda}, \citenamefont
  {Ishii}, \citenamefont {Yoshida}, \citenamefont {Mizuki}, \citenamefont
  {Kadono}, \citenamefont {Kumai}, \citenamefont {Kamiyama}, \citenamefont
  {Otomo}, \citenamefont {Murakami}, \citenamefont {Matsuishi},\ and\
  \citenamefont {Hosono}}]{Hiraishi2014}%
  \BibitemOpen
  \bibfield  {author} {\bibinfo {author} {\bibfnamefont {M.}~\bibnamefont
  {Hiraishi}}, \bibinfo {author} {\bibfnamefont {S.}~\bibnamefont {Iimura}},
  \bibinfo {author} {\bibfnamefont {K.~M.}\ \bibnamefont {Kojima}}, \bibinfo
  {author} {\bibfnamefont {J.}~\bibnamefont {Yamaura}}, \bibinfo {author}
  {\bibfnamefont {H.}~\bibnamefont {Hiraka}}, \bibinfo {author} {\bibfnamefont
  {K.}~\bibnamefont {Ikeda}}, \bibinfo {author} {\bibfnamefont
  {P.}~\bibnamefont {Miao}}, \bibinfo {author} {\bibfnamefont {Y.}~\bibnamefont
  {Ishikawa}}, \bibinfo {author} {\bibfnamefont {S.}~\bibnamefont {Torii}},
  \bibinfo {author} {\bibfnamefont {M.}~\bibnamefont {Miyazaki}}, \bibinfo
  {author} {\bibfnamefont {I.}~\bibnamefont {Yamauchi}}, \bibinfo {author}
  {\bibfnamefont {A.}~\bibnamefont {Koda}}, \bibinfo {author} {\bibfnamefont
  {K.}~\bibnamefont {Ishii}}, \bibinfo {author} {\bibfnamefont
  {M.}~\bibnamefont {Yoshida}}, \bibinfo {author} {\bibfnamefont
  {J.}~\bibnamefont {Mizuki}}, \bibinfo {author} {\bibfnamefont
  {R.}~\bibnamefont {Kadono}}, \bibinfo {author} {\bibfnamefont
  {R.}~\bibnamefont {Kumai}}, \bibinfo {author} {\bibfnamefont
  {T.}~\bibnamefont {Kamiyama}}, \bibinfo {author} {\bibfnamefont
  {T.}~\bibnamefont {Otomo}}, \bibinfo {author} {\bibfnamefont
  {Y.}~\bibnamefont {Murakami}}, \bibinfo {author} {\bibfnamefont
  {S.}~\bibnamefont {Matsuishi}}, \ and\ \bibinfo {author} {\bibfnamefont
  {H.}~\bibnamefont {Hosono}},\ }\href {\doibase 10.1038/nphys2906} {\bibfield
  {journal} {\bibinfo  {journal} {Nat. Phys.}\ }\textbf {\bibinfo {volume}
  {10}},\ \bibinfo {pages} {300} (\bibinfo {year} {2014})}\BibitemShut
  {NoStop}%
\bibitem [{\citenamefont {Iimura}\ \emph {et~al.}(2012)\citenamefont {Iimura},
  \citenamefont {Matuishi}, \citenamefont {Sato}, \citenamefont {Hanna},
  \citenamefont {Muraba}, \citenamefont {Kim}, \citenamefont {Kim},
  \citenamefont {Takata},\ and\ \citenamefont {Hosono}}]{Iimura2012}%
  \BibitemOpen
  \bibfield  {author} {\bibinfo {author} {\bibfnamefont {S.}~\bibnamefont
  {Iimura}}, \bibinfo {author} {\bibfnamefont {S.}~\bibnamefont {Matuishi}},
  \bibinfo {author} {\bibfnamefont {H.}~\bibnamefont {Sato}}, \bibinfo {author}
  {\bibfnamefont {T.}~\bibnamefont {Hanna}}, \bibinfo {author} {\bibfnamefont
  {Y.}~\bibnamefont {Muraba}}, \bibinfo {author} {\bibfnamefont {S.~W.}\
  \bibnamefont {Kim}}, \bibinfo {author} {\bibfnamefont {J.~E.}\ \bibnamefont
  {Kim}}, \bibinfo {author} {\bibfnamefont {M.}~\bibnamefont {Takata}}, \ and\
  \bibinfo {author} {\bibfnamefont {H.}~\bibnamefont {Hosono}},\ }\href
  {\doibase 10.1038/ncomms1913} {\bibfield  {journal} {\bibinfo  {journal}
  {Nature Communications}\ }\textbf {\bibinfo {volume} {3}},\ \bibinfo {pages}
  {943} (\bibinfo {year} {2012})}\BibitemShut {NoStop}%
\bibitem [{\citenamefont {Katayama}\ \emph {et~al.}(2013)\citenamefont
  {Katayama}, \citenamefont {Kudo}, \citenamefont {Onari}, \citenamefont
  {Mizukami}, \citenamefont {Sugawara}, \citenamefont {Sugiyama}, \citenamefont
  {Kitahama}, \citenamefont {Iba}, \citenamefont {Fujimura}, \citenamefont
  {Nishimoto}, \citenamefont {Nohara},\ and\ \citenamefont
  {Sawa}}]{Katayama2013}%
  \BibitemOpen
  \bibfield  {author} {\bibinfo {author} {\bibfnamefont {N.}~\bibnamefont
  {Katayama}}, \bibinfo {author} {\bibfnamefont {K.}~\bibnamefont {Kudo}},
  \bibinfo {author} {\bibfnamefont {S.}~\bibnamefont {Onari}}, \bibinfo
  {author} {\bibfnamefont {T.}~\bibnamefont {Mizukami}}, \bibinfo {author}
  {\bibfnamefont {K.}~\bibnamefont {Sugawara}}, \bibinfo {author}
  {\bibfnamefont {Y.}~\bibnamefont {Sugiyama}}, \bibinfo {author}
  {\bibfnamefont {Y.}~\bibnamefont {Kitahama}}, \bibinfo {author}
  {\bibfnamefont {K.}~\bibnamefont {Iba}}, \bibinfo {author} {\bibfnamefont
  {K.}~\bibnamefont {Fujimura}}, \bibinfo {author} {\bibfnamefont
  {N.}~\bibnamefont {Nishimoto}}, \bibinfo {author} {\bibfnamefont
  {M.}~\bibnamefont {Nohara}}, \ and\ \bibinfo {author} {\bibfnamefont
  {H.}~\bibnamefont {Sawa}},\ }\href {\doibase 10.7566/JPSJ.82.123702}
  {\bibfield  {journal} {\bibinfo  {journal} {J. Phys. Soc. Jpn.}\ }\textbf
  {\bibinfo {volume} {82}},\ \bibinfo {pages} {123702} (\bibinfo {year}
  {2013})}\BibitemShut {NoStop}%
\bibitem [{\citenamefont {Li}\ \emph {et~al.}(2015)\citenamefont {Li},
  \citenamefont {Liu}, \citenamefont {Zhou}, \citenamefont {Yang},
  \citenamefont {Shen}, \citenamefont {Li}, \citenamefont {Jiang},
  \citenamefont {Niu}, \citenamefont {Xie}, \citenamefont {Sun}, \citenamefont
  {Fan}, \citenamefont {Yao}, \citenamefont {Liu}, \citenamefont {Shi},\ and\
  \citenamefont {Xie}}]{Li2015}%
  \BibitemOpen
  \bibfield  {author} {\bibinfo {author} {\bibfnamefont {M.~Y.}\ \bibnamefont
  {Li}}, \bibinfo {author} {\bibfnamefont {Z.~T.}\ \bibnamefont {Liu}},
  \bibinfo {author} {\bibfnamefont {W.}~\bibnamefont {Zhou}}, \bibinfo {author}
  {\bibfnamefont {H.~F.}\ \bibnamefont {Yang}}, \bibinfo {author}
  {\bibfnamefont {D.~W.}\ \bibnamefont {Shen}}, \bibinfo {author}
  {\bibfnamefont {W.}~\bibnamefont {Li}}, \bibinfo {author} {\bibfnamefont
  {J.}~\bibnamefont {Jiang}}, \bibinfo {author} {\bibfnamefont {X.~H.}\
  \bibnamefont {Niu}}, \bibinfo {author} {\bibfnamefont {B.~P.}\ \bibnamefont
  {Xie}}, \bibinfo {author} {\bibfnamefont {Y.}~\bibnamefont {Sun}}, \bibinfo
  {author} {\bibfnamefont {C.~C.}\ \bibnamefont {Fan}}, \bibinfo {author}
  {\bibfnamefont {Q.}~\bibnamefont {Yao}}, \bibinfo {author} {\bibfnamefont
  {J.~S.}\ \bibnamefont {Liu}}, \bibinfo {author} {\bibfnamefont {Z.~X.}\
  \bibnamefont {Shi}}, \ and\ \bibinfo {author} {\bibfnamefont {X.~M.}\
  \bibnamefont {Xie}},\ }\href {\doibase 10.1103/PhysRevB.91.045112} {\bibfield
   {journal} {\bibinfo  {journal} {Phys. Rev. B}\ }\textbf {\bibinfo {volume}
  {91}},\ \bibinfo {pages} {045112} (\bibinfo {year} {2015})}\BibitemShut
  {NoStop}%
\bibitem [{\citenamefont {Kawasaki}\ \emph {et~al.}(2015)\citenamefont
  {Kawasaki}, \citenamefont {Mabuchi}, \citenamefont {Maeda}, \citenamefont
  {Adachi}, \citenamefont {Mizukami}, \citenamefont {Kudo}, \citenamefont
  {Nohara},\ and\ \citenamefont {Zheng}}]{Kawasaki2015}%
  \BibitemOpen
  \bibfield  {author} {\bibinfo {author} {\bibfnamefont {S.}~\bibnamefont
  {Kawasaki}}, \bibinfo {author} {\bibfnamefont {T.}~\bibnamefont {Mabuchi}},
  \bibinfo {author} {\bibfnamefont {S.}~\bibnamefont {Maeda}}, \bibinfo
  {author} {\bibfnamefont {T.}~\bibnamefont {Adachi}}, \bibinfo {author}
  {\bibfnamefont {T.}~\bibnamefont {Mizukami}}, \bibinfo {author}
  {\bibfnamefont {K.}~\bibnamefont {Kudo}}, \bibinfo {author} {\bibfnamefont
  {M.}~\bibnamefont {Nohara}}, \ and\ \bibinfo {author} {\bibfnamefont {G.-q.}\
  \bibnamefont {Zheng}},\ }\href {\doibase 10.1103/PhysRevB.92.180508}
  {\bibfield  {journal} {\bibinfo  {journal} {Phys. Rev. B}\ }\textbf {\bibinfo
  {volume} {92}},\ \bibinfo {pages} {180508} (\bibinfo {year}
  {2015})}\BibitemShut {NoStop}%
\bibitem [{\citenamefont {Jiang}\ \emph
  {et~al.}(2016{\natexlab{a}})\citenamefont {Jiang}, \citenamefont {Liu},
  \citenamefont {Cao}, \citenamefont {Birol}, \citenamefont {Allred},
  \citenamefont {Tian}, \citenamefont {Liu}, \citenamefont {Cho}, \citenamefont
  {Krogstad}, \citenamefont {Ma}, \citenamefont {Taddei}, \citenamefont
  {Tanatar}, \citenamefont {Hoesch}, \citenamefont {Prozorov}, \citenamefont
  {Rosenkranz}, \citenamefont {Uemura}, \citenamefont {Kotliar},\ and\
  \citenamefont {Ni}}]{Jiang2016}%
  \BibitemOpen
  \bibfield  {author} {\bibinfo {author} {\bibfnamefont {S.}~\bibnamefont
  {Jiang}}, \bibinfo {author} {\bibfnamefont {C.}~\bibnamefont {Liu}}, \bibinfo
  {author} {\bibfnamefont {H.}~\bibnamefont {Cao}}, \bibinfo {author}
  {\bibfnamefont {T.}~\bibnamefont {Birol}}, \bibinfo {author} {\bibfnamefont
  {J.~M.}\ \bibnamefont {Allred}}, \bibinfo {author} {\bibfnamefont
  {W.}~\bibnamefont {Tian}}, \bibinfo {author} {\bibfnamefont {L.}~\bibnamefont
  {Liu}}, \bibinfo {author} {\bibfnamefont {K.}~\bibnamefont {Cho}}, \bibinfo
  {author} {\bibfnamefont {M.~J.}\ \bibnamefont {Krogstad}}, \bibinfo {author}
  {\bibfnamefont {J.}~\bibnamefont {Ma}}, \bibinfo {author} {\bibfnamefont
  {K.~M.}\ \bibnamefont {Taddei}}, \bibinfo {author} {\bibfnamefont {M.~A.}\
  \bibnamefont {Tanatar}}, \bibinfo {author} {\bibfnamefont {M.}~\bibnamefont
  {Hoesch}}, \bibinfo {author} {\bibfnamefont {R.}~\bibnamefont {Prozorov}},
  \bibinfo {author} {\bibfnamefont {S.}~\bibnamefont {Rosenkranz}}, \bibinfo
  {author} {\bibfnamefont {Y.~J.}\ \bibnamefont {Uemura}}, \bibinfo {author}
  {\bibfnamefont {G.}~\bibnamefont {Kotliar}}, \ and\ \bibinfo {author}
  {\bibfnamefont {N.}~\bibnamefont {Ni}},\ }\href {\doibase
  10.1103/PhysRevB.93.054522} {\bibfield  {journal} {\bibinfo  {journal} {Phys.
  Rev. B}\ }\textbf {\bibinfo {volume} {93}},\ \bibinfo {pages} {054522}
  (\bibinfo {year} {2016}{\natexlab{a}})}\BibitemShut {NoStop}%
\bibitem [{\citenamefont {Wu}\ \emph {et~al.}(2010)\citenamefont {Wu},
  \citenamefont {Bari{\v{s}}i{\'{c}}}, \citenamefont {Kallina}, \citenamefont
  {Faridian}, \citenamefont {Gorshunov}, \citenamefont {Drichko}, \citenamefont
  {Li}, \citenamefont {Lin}, \citenamefont {Cao}, \citenamefont {Xu},
  \citenamefont {Wang},\ and\ \citenamefont {Dressel}}]{Wu2010}%
  \BibitemOpen
  \bibfield  {author} {\bibinfo {author} {\bibfnamefont {D.}~\bibnamefont
  {Wu}}, \bibinfo {author} {\bibfnamefont {N.}~\bibnamefont
  {Bari{\v{s}}i{\'{c}}}}, \bibinfo {author} {\bibfnamefont {P.}~\bibnamefont
  {Kallina}}, \bibinfo {author} {\bibfnamefont {A.}~\bibnamefont {Faridian}},
  \bibinfo {author} {\bibfnamefont {B.}~\bibnamefont {Gorshunov}}, \bibinfo
  {author} {\bibfnamefont {N.}~\bibnamefont {Drichko}}, \bibinfo {author}
  {\bibfnamefont {L.~J.}\ \bibnamefont {Li}}, \bibinfo {author} {\bibfnamefont
  {X.}~\bibnamefont {Lin}}, \bibinfo {author} {\bibfnamefont {G.~H.}\
  \bibnamefont {Cao}}, \bibinfo {author} {\bibfnamefont {Z.~A.}\ \bibnamefont
  {Xu}}, \bibinfo {author} {\bibfnamefont {N.~L.}\ \bibnamefont {Wang}}, \ and\
  \bibinfo {author} {\bibfnamefont {M.}~\bibnamefont {Dressel}},\ }\href
  {\doibase 10.1103/PhysRevB.81.100512} {\bibfield  {journal} {\bibinfo
  {journal} {Phys. Rev. B}\ }\textbf {\bibinfo {volume} {81}},\ \bibinfo
  {pages} {100512} (\bibinfo {year} {2010})}\BibitemShut {NoStop}%
\bibitem [{\citenamefont {Dai}\ \emph {et~al.}(2015)\citenamefont {Dai},
  \citenamefont {Miao}, \citenamefont {Xing}, \citenamefont {Wang},
  \citenamefont {Wang}, \citenamefont {Xiao}, \citenamefont {Qian},
  \citenamefont {Richard}, \citenamefont {Qiu}, \citenamefont {Yu},
  \citenamefont {Jin}, \citenamefont {Wang}, \citenamefont {Johnson},
  \citenamefont {Homes},\ and\ \citenamefont {Ding}}]{Dai2015}%
  \BibitemOpen
  \bibfield  {author} {\bibinfo {author} {\bibfnamefont {Y.~M.}\ \bibnamefont
  {Dai}}, \bibinfo {author} {\bibfnamefont {H.}~\bibnamefont {Miao}}, \bibinfo
  {author} {\bibfnamefont {L.~Y.}\ \bibnamefont {Xing}}, \bibinfo {author}
  {\bibfnamefont {X.~C.}\ \bibnamefont {Wang}}, \bibinfo {author}
  {\bibfnamefont {P.~S.}\ \bibnamefont {Wang}}, \bibinfo {author}
  {\bibfnamefont {H.}~\bibnamefont {Xiao}}, \bibinfo {author} {\bibfnamefont
  {T.}~\bibnamefont {Qian}}, \bibinfo {author} {\bibfnamefont {P.}~\bibnamefont
  {Richard}}, \bibinfo {author} {\bibfnamefont {X.~G.}\ \bibnamefont {Qiu}},
  \bibinfo {author} {\bibfnamefont {W.}~\bibnamefont {Yu}}, \bibinfo {author}
  {\bibfnamefont {C.~Q.}\ \bibnamefont {Jin}}, \bibinfo {author} {\bibfnamefont
  {Z.}~\bibnamefont {Wang}}, \bibinfo {author} {\bibfnamefont {P.~D.}\
  \bibnamefont {Johnson}}, \bibinfo {author} {\bibfnamefont {C.~C.}\
  \bibnamefont {Homes}}, \ and\ \bibinfo {author} {\bibfnamefont
  {H.}~\bibnamefont {Ding}},\ }\href {\doibase 10.1103/PhysRevX.5.031035}
  {\bibfield  {journal} {\bibinfo  {journal} {Phys. Rev. X}\ }\textbf {\bibinfo
  {volume} {5}},\ \bibinfo {pages} {031035} (\bibinfo {year}
  {2015})}\BibitemShut {NoStop}%
\bibitem [{\citenamefont {Kudo}\ \emph {et~al.}(2014)\citenamefont {Kudo},
  \citenamefont {Kitahama}, \citenamefont {Fujimura}, \citenamefont {Mizukami},
  \citenamefont {Ota},\ and\ \citenamefont {Nohara}}]{Kudo2014}%
  \BibitemOpen
  \bibfield  {author} {\bibinfo {author} {\bibfnamefont {K.}~\bibnamefont
  {Kudo}}, \bibinfo {author} {\bibfnamefont {Y.}~\bibnamefont {Kitahama}},
  \bibinfo {author} {\bibfnamefont {K.}~\bibnamefont {Fujimura}}, \bibinfo
  {author} {\bibfnamefont {T.}~\bibnamefont {Mizukami}}, \bibinfo {author}
  {\bibfnamefont {H.}~\bibnamefont {Ota}}, \ and\ \bibinfo {author}
  {\bibfnamefont {M.}~\bibnamefont {Nohara}},\ }\href {\doibase
  10.7566/JPSJ.83.093705} {\bibfield  {journal} {\bibinfo  {journal} {J. Phys.
  Soc. Jpn.}\ }\textbf {\bibinfo {volume} {83}},\ \bibinfo {pages} {093705}
  (\bibinfo {year} {2014})}\BibitemShut {NoStop}%
\bibitem [{\citenamefont {Homes}\ \emph {et~al.}(1993)\citenamefont {Homes},
  \citenamefont {Reedyk}, \citenamefont {Cradles},\ and\ \citenamefont
  {Timusk}}]{Homes1993}%
  \BibitemOpen
  \bibfield  {author} {\bibinfo {author} {\bibfnamefont {C.~C.}\ \bibnamefont
  {Homes}}, \bibinfo {author} {\bibfnamefont {M.}~\bibnamefont {Reedyk}},
  \bibinfo {author} {\bibfnamefont {D.~A.}\ \bibnamefont {Cradles}}, \ and\
  \bibinfo {author} {\bibfnamefont {T.}~\bibnamefont {Timusk}},\ }\href
  {\doibase 10.1364/AO.32.002976} {\bibfield  {journal} {\bibinfo  {journal}
  {Applied Optics}\ }\textbf {\bibinfo {volume} {32}},\ \bibinfo {pages} {2976}
  (\bibinfo {year} {1993})}\BibitemShut {NoStop}%
\bibitem [{\citenamefont {Dai}\ \emph {et~al.}(2013)\citenamefont {Dai},
  \citenamefont {Xu}, \citenamefont {Shen}, \citenamefont {Xiao}, \citenamefont
  {Wen}, \citenamefont {Qiu}, \citenamefont {Homes},\ and\ \citenamefont
  {Lobo}}]{Dai2013}%
  \BibitemOpen
  \bibfield  {author} {\bibinfo {author} {\bibfnamefont {Y.~M.}\ \bibnamefont
  {Dai}}, \bibinfo {author} {\bibfnamefont {B.}~\bibnamefont {Xu}}, \bibinfo
  {author} {\bibfnamefont {B.}~\bibnamefont {Shen}}, \bibinfo {author}
  {\bibfnamefont {H.}~\bibnamefont {Xiao}}, \bibinfo {author} {\bibfnamefont
  {H.~H.}\ \bibnamefont {Wen}}, \bibinfo {author} {\bibfnamefont {X.~G.}\
  \bibnamefont {Qiu}}, \bibinfo {author} {\bibfnamefont {C.~C.}\ \bibnamefont
  {Homes}}, \ and\ \bibinfo {author} {\bibfnamefont {R.~P. S.~M.}\ \bibnamefont
  {Lobo}},\ }\href {\doibase 10.1103/PhysRevLett.111.117001} {\bibfield
  {journal} {\bibinfo  {journal} {Physical Review Letters}\ }\textbf {\bibinfo
  {volume} {111}},\ \bibinfo {pages} {117001} (\bibinfo {year}
  {2013})}\BibitemShut {NoStop}%
\bibitem [{\citenamefont {Analytis}\ \emph {et~al.}(2014)\citenamefont
  {Analytis}, \citenamefont {Kuo}, \citenamefont {McDonald}, \citenamefont
  {Wartenbe}, \citenamefont {Rourke}, \citenamefont {Hussey},\ and\
  \citenamefont {Fisher}}]{Analytis2014}%
  \BibitemOpen
  \bibfield  {author} {\bibinfo {author} {\bibfnamefont {J.~G.}\ \bibnamefont
  {Analytis}}, \bibinfo {author} {\bibfnamefont {H.-H.}\ \bibnamefont {Kuo}},
  \bibinfo {author} {\bibfnamefont {R.~D.}\ \bibnamefont {McDonald}}, \bibinfo
  {author} {\bibfnamefont {M.}~\bibnamefont {Wartenbe}}, \bibinfo {author}
  {\bibfnamefont {P.~M.~C.}\ \bibnamefont {Rourke}}, \bibinfo {author}
  {\bibfnamefont {N.~E.}\ \bibnamefont {Hussey}}, \ and\ \bibinfo {author}
  {\bibfnamefont {I.~R.}\ \bibnamefont {Fisher}},\ }\href {\doibase
  10.1038/nphys2869} {\bibfield  {journal} {\bibinfo  {journal} {Nat. Phys.}\
  }\textbf {\bibinfo {volume} {10}},\ \bibinfo {pages} {194} (\bibinfo {year}
  {2014})}\BibitemShut {NoStop}%
\bibitem [{\citenamefont {Hu}\ \emph {et~al.}(2008)\citenamefont {Hu},
  \citenamefont {Dong}, \citenamefont {Li}, \citenamefont {Li}, \citenamefont
  {Zheng}, \citenamefont {Chen}, \citenamefont {Luo},\ and\ \citenamefont
  {Wang}}]{Hu2008}%
  \BibitemOpen
  \bibfield  {author} {\bibinfo {author} {\bibfnamefont {W.~Z.}\ \bibnamefont
  {Hu}}, \bibinfo {author} {\bibfnamefont {J.}~\bibnamefont {Dong}}, \bibinfo
  {author} {\bibfnamefont {G.}~\bibnamefont {Li}}, \bibinfo {author}
  {\bibfnamefont {Z.}~\bibnamefont {Li}}, \bibinfo {author} {\bibfnamefont
  {P.}~\bibnamefont {Zheng}}, \bibinfo {author} {\bibfnamefont {G.~F.}\
  \bibnamefont {Chen}}, \bibinfo {author} {\bibfnamefont {J.~L.}\ \bibnamefont
  {Luo}}, \ and\ \bibinfo {author} {\bibfnamefont {N.~L.}\ \bibnamefont
  {Wang}},\ }\href {\doibase 10.1103/PhysRevLett.101.257005} {\bibfield
  {journal} {\bibinfo  {journal} {Phys. Rev. Lett.}\ }\textbf {\bibinfo
  {volume} {101}},\ \bibinfo {pages} {257005} (\bibinfo {year}
  {2008})}\BibitemShut {NoStop}%
\bibitem [{\citenamefont {Homes}\ \emph {et~al.}(2015)\citenamefont {Homes},
  \citenamefont {Dai}, \citenamefont {Wen}, \citenamefont {Xu},\ and\
  \citenamefont {Gu}}]{Homes2015}%
  \BibitemOpen
  \bibfield  {author} {\bibinfo {author} {\bibfnamefont {C.~C.}\ \bibnamefont
  {Homes}}, \bibinfo {author} {\bibfnamefont {Y.~M.}\ \bibnamefont {Dai}},
  \bibinfo {author} {\bibfnamefont {J.~S.}\ \bibnamefont {Wen}}, \bibinfo
  {author} {\bibfnamefont {Z.~J.}\ \bibnamefont {Xu}}, \ and\ \bibinfo {author}
  {\bibfnamefont {G.~D.}\ \bibnamefont {Gu}},\ }\href {\doibase
  10.1103/PhysRevB.91.144503} {\bibfield  {journal} {\bibinfo  {journal} {Phys.
  Rev. B}\ }\textbf {\bibinfo {volume} {91}},\ \bibinfo {pages} {144503}
  (\bibinfo {year} {2015})}\BibitemShut {NoStop}%
\bibitem [{\citenamefont {Nakajima}\ \emph
  {et~al.}(2014{\natexlab{a}})\citenamefont {Nakajima}, \citenamefont {Ishida},
  \citenamefont {Tanaka}, \citenamefont {Kihou}, \citenamefont {Tomioka},
  \citenamefont {Saito}, \citenamefont {Lee}, \citenamefont {Fukazawa},
  \citenamefont {Kohori}, \citenamefont {Kakeshita}, \citenamefont {Iyo},
  \citenamefont {Ito}, \citenamefont {Eisaki},\ and\ \citenamefont
  {Uchida}}]{Nakajima2014}%
  \BibitemOpen
  \bibfield  {author} {\bibinfo {author} {\bibfnamefont {M.}~\bibnamefont
  {Nakajima}}, \bibinfo {author} {\bibfnamefont {S.}~\bibnamefont {Ishida}},
  \bibinfo {author} {\bibfnamefont {T.}~\bibnamefont {Tanaka}}, \bibinfo
  {author} {\bibfnamefont {K.}~\bibnamefont {Kihou}}, \bibinfo {author}
  {\bibfnamefont {Y.}~\bibnamefont {Tomioka}}, \bibinfo {author} {\bibfnamefont
  {T.}~\bibnamefont {Saito}}, \bibinfo {author} {\bibfnamefont {C.-H.}\
  \bibnamefont {Lee}}, \bibinfo {author} {\bibfnamefont {H.}~\bibnamefont
  {Fukazawa}}, \bibinfo {author} {\bibfnamefont {Y.}~\bibnamefont {Kohori}},
  \bibinfo {author} {\bibfnamefont {T.}~\bibnamefont {Kakeshita}}, \bibinfo
  {author} {\bibfnamefont {A.}~\bibnamefont {Iyo}}, \bibinfo {author}
  {\bibfnamefont {T.}~\bibnamefont {Ito}}, \bibinfo {author} {\bibfnamefont
  {H.}~\bibnamefont {Eisaki}}, \ and\ \bibinfo {author} {\bibfnamefont {S.-i.}\
  \bibnamefont {Uchida}},\ }\href {\doibase 10.7566/JPSJ.83.104703} {\bibfield
  {journal} {\bibinfo  {journal} {J. Phys. Soc. Jpn.}\ }\textbf {\bibinfo
  {volume} {83}},\ \bibinfo {pages} {104703} (\bibinfo {year}
  {2014}{\natexlab{a}})}\BibitemShut {NoStop}%
\bibitem [{\citenamefont {Nakajima}\ \emph
  {et~al.}(2014{\natexlab{b}})\citenamefont {Nakajima}, \citenamefont {Ishida},
  \citenamefont {Tanaka}, \citenamefont {Kihou}, \citenamefont {Tomioka},
  \citenamefont {Saito}, \citenamefont {Lee}, \citenamefont {Fukazawa},
  \citenamefont {Kohori}, \citenamefont {Kakeshita}, \citenamefont {Iyo},
  \citenamefont {Ito}, \citenamefont {Eisaki},\ and\ \citenamefont
  {Uchida}}]{Nakajima2014a}%
  \BibitemOpen
  \bibfield  {author} {\bibinfo {author} {\bibfnamefont {M.}~\bibnamefont
  {Nakajima}}, \bibinfo {author} {\bibfnamefont {S.}~\bibnamefont {Ishida}},
  \bibinfo {author} {\bibfnamefont {T.}~\bibnamefont {Tanaka}}, \bibinfo
  {author} {\bibfnamefont {K.}~\bibnamefont {Kihou}}, \bibinfo {author}
  {\bibfnamefont {Y.}~\bibnamefont {Tomioka}}, \bibinfo {author} {\bibfnamefont
  {T.}~\bibnamefont {Saito}}, \bibinfo {author} {\bibfnamefont {C.~H.}\
  \bibnamefont {Lee}}, \bibinfo {author} {\bibfnamefont {H.}~\bibnamefont
  {Fukazawa}}, \bibinfo {author} {\bibfnamefont {Y.}~\bibnamefont {Kohori}},
  \bibinfo {author} {\bibfnamefont {T.}~\bibnamefont {Kakeshita}}, \bibinfo
  {author} {\bibfnamefont {A.}~\bibnamefont {Iyo}}, \bibinfo {author}
  {\bibfnamefont {T.}~\bibnamefont {Ito}}, \bibinfo {author} {\bibfnamefont
  {H.}~\bibnamefont {Eisaki}}, \ and\ \bibinfo {author} {\bibfnamefont
  {S.}~\bibnamefont {Uchida}},\ }\href {\doibase 10.1038/srep05873} {\bibfield
  {journal} {\bibinfo  {journal} {Sci. Rep.}\ }\textbf {\bibinfo {volume} {4}}
  (\bibinfo {year} {2014}{\natexlab{b}}),\ 10.1038/srep05873}\BibitemShut
  {NoStop}%
\bibitem [{\citenamefont {Wang}\ \emph {et~al.}(2012)\citenamefont {Wang},
  \citenamefont {Hu}, \citenamefont {Chen}, \citenamefont {Yuan}, \citenamefont
  {Li}, \citenamefont {Chen},\ and\ \citenamefont {Xiang}}]{Wang2012}%
  \BibitemOpen
  \bibfield  {author} {\bibinfo {author} {\bibfnamefont {N.~L.}\ \bibnamefont
  {Wang}}, \bibinfo {author} {\bibfnamefont {W.~Z.}\ \bibnamefont {Hu}},
  \bibinfo {author} {\bibfnamefont {Z.~G.}\ \bibnamefont {Chen}}, \bibinfo
  {author} {\bibfnamefont {R.~H.}\ \bibnamefont {Yuan}}, \bibinfo {author}
  {\bibfnamefont {G.}~\bibnamefont {Li}}, \bibinfo {author} {\bibfnamefont
  {G.~F.}\ \bibnamefont {Chen}}, \ and\ \bibinfo {author} {\bibfnamefont
  {T.}~\bibnamefont {Xiang}},\ }\href {\doibase 10.1088/0953-8984/24/29/294202}
  {\bibfield  {journal} {\bibinfo  {journal} {Journal of Physics: Condensed
  Matter}\ }\textbf {\bibinfo {volume} {24}},\ \bibinfo {pages} {294202}
  (\bibinfo {year} {2012})}\BibitemShut {NoStop}%
\bibitem [{\citenamefont {Richard}\ \emph {et~al.}(2011)\citenamefont
  {Richard}, \citenamefont {Sato}, \citenamefont {Nakayama}, \citenamefont
  {Takahashi},\ and\ \citenamefont {Ding}}]{Richard2011}%
  \BibitemOpen
  \bibfield  {author} {\bibinfo {author} {\bibfnamefont {P.}~\bibnamefont
  {Richard}}, \bibinfo {author} {\bibfnamefont {T.}~\bibnamefont {Sato}},
  \bibinfo {author} {\bibfnamefont {K.}~\bibnamefont {Nakayama}}, \bibinfo
  {author} {\bibfnamefont {T.}~\bibnamefont {Takahashi}}, \ and\ \bibinfo
  {author} {\bibfnamefont {H.}~\bibnamefont {Ding}},\ }\href {\doibase
  10.1088/0034-4885/74/12/124512} {\bibfield  {journal} {\bibinfo  {journal}
  {Reports on Progress in Physics}\ }\textbf {\bibinfo {volume} {74}},\
  \bibinfo {pages} {124512} (\bibinfo {year} {2011})}\BibitemShut {NoStop}%
\bibitem [{\citenamefont {Hancock}\ \emph {et~al.}(2010)\citenamefont
  {Hancock}, \citenamefont {Mirzaei}, \citenamefont {Gillett}, \citenamefont
  {Sebastian}, \citenamefont {Teyssier}, \citenamefont {Viennois},
  \citenamefont {Giannini},\ and\ \citenamefont {van~der Marel}}]{Hancock2010}%
  \BibitemOpen
  \bibfield  {author} {\bibinfo {author} {\bibfnamefont {J.~N.}\ \bibnamefont
  {Hancock}}, \bibinfo {author} {\bibfnamefont {S.~I.}\ \bibnamefont
  {Mirzaei}}, \bibinfo {author} {\bibfnamefont {J.}~\bibnamefont {Gillett}},
  \bibinfo {author} {\bibfnamefont {S.~E.}\ \bibnamefont {Sebastian}}, \bibinfo
  {author} {\bibfnamefont {J.}~\bibnamefont {Teyssier}}, \bibinfo {author}
  {\bibfnamefont {R.}~\bibnamefont {Viennois}}, \bibinfo {author}
  {\bibfnamefont {E.}~\bibnamefont {Giannini}}, \ and\ \bibinfo {author}
  {\bibfnamefont {D.}~\bibnamefont {van~der Marel}},\ }\href {\doibase
  10.1103/PhysRevB.82.014523} {\bibfield  {journal} {\bibinfo  {journal} {Phys.
  Rev. B}\ }\textbf {\bibinfo {volume} {82}},\ \bibinfo {pages} {014523}
  (\bibinfo {year} {2010})}\BibitemShut {NoStop}%
\bibitem [{\citenamefont {Dai}\ \emph {et~al.}(2012)\citenamefont {Dai},
  \citenamefont {Hu},\ and\ \citenamefont {Dagotto}}]{Dai2012}%
  \BibitemOpen
  \bibfield  {author} {\bibinfo {author} {\bibfnamefont {P.}~\bibnamefont
  {Dai}}, \bibinfo {author} {\bibfnamefont {J.}~\bibnamefont {Hu}}, \ and\
  \bibinfo {author} {\bibfnamefont {E.}~\bibnamefont {Dagotto}},\ }\href
  {\doibase 10.1038/nphys2438} {\bibfield  {journal} {\bibinfo  {journal} {Nat.
  Phys.}\ }\textbf {\bibinfo {volume} {8}},\ \bibinfo {pages} {709} (\bibinfo
  {year} {2012})}\BibitemShut {NoStop}%
\bibitem [{\citenamefont {Ikeda}\ \emph {et~al.}(2010)\citenamefont {Ikeda},
  \citenamefont {Arita},\ and\ \citenamefont {Kune{\v{s}}}}]{Ikeda2010}%
  \BibitemOpen
  \bibfield  {author} {\bibinfo {author} {\bibfnamefont {H.}~\bibnamefont
  {Ikeda}}, \bibinfo {author} {\bibfnamefont {R.}~\bibnamefont {Arita}}, \ and\
  \bibinfo {author} {\bibfnamefont {J.}~\bibnamefont {Kune{\v{s}}}},\ }\href
  {\doibase 10.1103/PhysRevB.82.024508} {\bibfield  {journal} {\bibinfo
  {journal} {Phys. Rev. B}\ }\textbf {\bibinfo {volume} {82}},\ \bibinfo
  {pages} {024508} (\bibinfo {year} {2010})}\BibitemShut {NoStop}%
\bibitem [{\citenamefont {Diehl}\ \emph {et~al.}(2014)\citenamefont {Diehl},
  \citenamefont {Backes}, \citenamefont {Guterding}, \citenamefont {Jeschke},\
  and\ \citenamefont {Valent{\'{\i}}}}]{diehl}%
  \BibitemOpen
  \bibfield  {author} {\bibinfo {author} {\bibfnamefont {J.}~\bibnamefont
  {Diehl}}, \bibinfo {author} {\bibfnamefont {S.}~\bibnamefont {Backes}},
  \bibinfo {author} {\bibfnamefont {D.}~\bibnamefont {Guterding}}, \bibinfo
  {author} {\bibfnamefont {H.~O.}\ \bibnamefont {Jeschke}}, \ and\ \bibinfo
  {author} {\bibfnamefont {R.}~\bibnamefont {Valent{\'{\i}}}},\ }\href
  {\doibase 10.1103/PhysRevB.90.085110} {\bibfield  {journal} {\bibinfo
  {journal} {Phys. Rev. B}\ }\textbf {\bibinfo {volume} {90}},\ \bibinfo
  {pages} {085110} (\bibinfo {year} {2014})}\BibitemShut {NoStop}%
\bibitem [{\citenamefont {Zhang}\ \emph {et~al.}(2014)\citenamefont {Zhang},
  \citenamefont {Harriger}, \citenamefont {Yin}, \citenamefont {Lv},
  \citenamefont {Wang}, \citenamefont {Tan}, \citenamefont {Song},
  \citenamefont {Abernathy}, \citenamefont {Tian}, \citenamefont {Egami},
  \citenamefont {Haule}, \citenamefont {Kotliar},\ and\ \citenamefont
  {Dai}}]{Zhang2014}%
  \BibitemOpen
  \bibfield  {author} {\bibinfo {author} {\bibfnamefont {C.}~\bibnamefont
  {Zhang}}, \bibinfo {author} {\bibfnamefont {L.~W.}\ \bibnamefont {Harriger}},
  \bibinfo {author} {\bibfnamefont {Z.}~\bibnamefont {Yin}}, \bibinfo {author}
  {\bibfnamefont {W.}~\bibnamefont {Lv}}, \bibinfo {author} {\bibfnamefont
  {M.}~\bibnamefont {Wang}}, \bibinfo {author} {\bibfnamefont {G.}~\bibnamefont
  {Tan}}, \bibinfo {author} {\bibfnamefont {Y.}~\bibnamefont {Song}}, \bibinfo
  {author} {\bibfnamefont {D.~L.}\ \bibnamefont {Abernathy}}, \bibinfo {author}
  {\bibfnamefont {W.}~\bibnamefont {Tian}}, \bibinfo {author} {\bibfnamefont
  {T.}~\bibnamefont {Egami}}, \bibinfo {author} {\bibfnamefont
  {K.}~\bibnamefont {Haule}}, \bibinfo {author} {\bibfnamefont
  {G.}~\bibnamefont {Kotliar}}, \ and\ \bibinfo {author} {\bibfnamefont
  {P.}~\bibnamefont {Dai}},\ }\href {\doibase 10.1103/PhysRevLett.112.217202}
  {\bibfield  {journal} {\bibinfo  {journal} {Physical Review Letters}\
  }\textbf {\bibinfo {volume} {112}},\ \bibinfo {pages} {217202} (\bibinfo
  {year} {2014})}\BibitemShut {NoStop}%
\bibitem [{\citenamefont {Si}\ \emph {et~al.}(2009)\citenamefont {Si},
  \citenamefont {Abrahams}, \citenamefont {Dai},\ and\ \citenamefont
  {Zhu}}]{Si2009}%
  \BibitemOpen
  \bibfield  {author} {\bibinfo {author} {\bibfnamefont {Q.}~\bibnamefont
  {Si}}, \bibinfo {author} {\bibfnamefont {E.}~\bibnamefont {Abrahams}},
  \bibinfo {author} {\bibfnamefont {J.}~\bibnamefont {Dai}}, \ and\ \bibinfo
  {author} {\bibfnamefont {J.-X.}\ \bibnamefont {Zhu}},\ }\href {\doibase
  10.1088/1367-2630/11/4/045001} {\bibfield  {journal} {\bibinfo  {journal}
  {New J. Phys.}\ }\textbf {\bibinfo {volume} {11}},\ \bibinfo {pages} {045001}
  (\bibinfo {year} {2009})}\BibitemShut {NoStop}%
\bibitem [{\citenamefont {de¡¯ Medici}\ \emph {et~al.}(2009)\citenamefont {de¡¯
  Medici}, \citenamefont {Hassan}, \citenamefont {Capone},\ and\ \citenamefont
  {Dai}}]{DeMedici2009}%
  \BibitemOpen
  \bibfield  {author} {\bibinfo {author} {\bibfnamefont {L.}~\bibnamefont {de¡¯
  Medici}}, \bibinfo {author} {\bibfnamefont {S.~R.}\ \bibnamefont {Hassan}},
  \bibinfo {author} {\bibfnamefont {M.}~\bibnamefont {Capone}}, \ and\ \bibinfo
  {author} {\bibfnamefont {X.}~\bibnamefont {Dai}},\ }\href {\doibase
  10.1103/PhysRevLett.102.126401} {\bibfield  {journal} {\bibinfo  {journal}
  {Phys. Revs Lett.}\ }\textbf {\bibinfo {volume} {102}},\ \bibinfo {pages}
  {126401} (\bibinfo {year} {2009})}\BibitemShut {NoStop}%
\bibitem [{\citenamefont {Neupane}\ \emph {et~al.}(2009)\citenamefont
  {Neupane}, \citenamefont {Richard}, \citenamefont {Pan}, \citenamefont {Xu},
  \citenamefont {Jin}, \citenamefont {Mandrus}, \citenamefont {Dai},
  \citenamefont {Fang}, \citenamefont {Wang},\ and\ \citenamefont
  {Ding}}]{Neupane2009}%
  \BibitemOpen
  \bibfield  {author} {\bibinfo {author} {\bibfnamefont {M.}~\bibnamefont
  {Neupane}}, \bibinfo {author} {\bibfnamefont {P.}~\bibnamefont {Richard}},
  \bibinfo {author} {\bibfnamefont {Z.-H.}\ \bibnamefont {Pan}}, \bibinfo
  {author} {\bibfnamefont {Y.-M.}\ \bibnamefont {Xu}}, \bibinfo {author}
  {\bibfnamefont {R.}~\bibnamefont {Jin}}, \bibinfo {author} {\bibfnamefont
  {D.}~\bibnamefont {Mandrus}}, \bibinfo {author} {\bibfnamefont
  {X.}~\bibnamefont {Dai}}, \bibinfo {author} {\bibfnamefont {Z.}~\bibnamefont
  {Fang}}, \bibinfo {author} {\bibfnamefont {Z.}~\bibnamefont {Wang}}, \ and\
  \bibinfo {author} {\bibfnamefont {H.}~\bibnamefont {Ding}},\ }\href {\doibase
  10.1103/PhysRevLett.103.097001} {\bibfield  {journal} {\bibinfo  {journal}
  {Phys. Revs Lett.}\ }\textbf {\bibinfo {volume} {103}},\ \bibinfo {pages}
  {097001} (\bibinfo {year} {2009})}\BibitemShut {NoStop}%
\bibitem [{\citenamefont {Jiang}\ \emph {et~al.}(2015)\citenamefont {Jiang},
  \citenamefont {Liu}, \citenamefont {Cao}, \citenamefont {Birol},
  \citenamefont {Allred}, \citenamefont {Tian}, \citenamefont {Liu},
  \citenamefont {Cho}, \citenamefont {Krogstad}, \citenamefont {Ma},
  \citenamefont {Taddei}, \citenamefont {Tanatar}, \citenamefont {Hoesch},
  \citenamefont {Prozorov}, \citenamefont {Rosenkranz}, \citenamefont {Uemura},
  \citenamefont {Kotliar},\ and\ \citenamefont {Ni}}]{Jiang2015}%
  \BibitemOpen
  \bibfield  {author} {\bibinfo {author} {\bibfnamefont {S.}~\bibnamefont
  {Jiang}}, \bibinfo {author} {\bibfnamefont {C.}~\bibnamefont {Liu}}, \bibinfo
  {author} {\bibfnamefont {H.}~\bibnamefont {Cao}}, \bibinfo {author}
  {\bibfnamefont {T.}~\bibnamefont {Birol}}, \bibinfo {author} {\bibfnamefont
  {J.~M.}\ \bibnamefont {Allred}}, \bibinfo {author} {\bibfnamefont
  {W.}~\bibnamefont {Tian}}, \bibinfo {author} {\bibfnamefont {L.}~\bibnamefont
  {Liu}}, \bibinfo {author} {\bibfnamefont {K.}~\bibnamefont {Cho}}, \bibinfo
  {author} {\bibfnamefont {M.~J.}\ \bibnamefont {Krogstad}}, \bibinfo {author}
  {\bibfnamefont {J.}~\bibnamefont {Ma}}, \bibinfo {author} {\bibfnamefont
  {K.~M.}\ \bibnamefont {Taddei}}, \bibinfo {author} {\bibfnamefont {M.~A.}\
  \bibnamefont {Tanatar}}, \bibinfo {author} {\bibfnamefont {M.}~\bibnamefont
  {Hoesch}}, \bibinfo {author} {\bibfnamefont {R.}~\bibnamefont {Prozorov}},
  \bibinfo {author} {\bibfnamefont {S.}~\bibnamefont {Rosenkranz}}, \bibinfo
  {author} {\bibfnamefont {Y.~J.}\ \bibnamefont {Uemura}}, \bibinfo {author}
  {\bibfnamefont {G.}~\bibnamefont {Kotliar}}, \ and\ \bibinfo {author}
  {\bibfnamefont {N.}~\bibnamefont {Ni}},\ }\href
  {http://arxiv.org/abs/1505.05881} {\  (\bibinfo {year} {2015})},\ \Eprint
  {http://arxiv.org/abs/1505.05881} {arXiv:1505.05881} \BibitemShut {NoStop}%
\bibitem [{\citenamefont {Jiang}\ \emph
  {et~al.}(2016{\natexlab{b}})\citenamefont {Jiang}, \citenamefont {Liu},
  \citenamefont {Schutt}, \citenamefont {Hallas}, \citenamefont {Shen},
  \citenamefont {Tian}, \citenamefont {Emmanouilidou}, \citenamefont {Shi},
  \citenamefont {Luke}, \citenamefont {Uemura}, \citenamefont {Fernandes},\
  and\ \citenamefont {Ni}}]{Jiang2016a}%
  \BibitemOpen
  \bibfield  {author} {\bibinfo {author} {\bibfnamefont {S.}~\bibnamefont
  {Jiang}}, \bibinfo {author} {\bibfnamefont {L.}~\bibnamefont {Liu}}, \bibinfo
  {author} {\bibfnamefont {M.}~\bibnamefont {Schutt}}, \bibinfo {author}
  {\bibfnamefont {A.~M.}\ \bibnamefont {Hallas}}, \bibinfo {author}
  {\bibfnamefont {B.}~\bibnamefont {Shen}}, \bibinfo {author} {\bibfnamefont
  {W.}~\bibnamefont {Tian}}, \bibinfo {author} {\bibfnamefont {E.}~\bibnamefont
  {Emmanouilidou}}, \bibinfo {author} {\bibfnamefont {A.}~\bibnamefont {Shi}},
  \bibinfo {author} {\bibfnamefont {G.~M.}\ \bibnamefont {Luke}}, \bibinfo
  {author} {\bibfnamefont {Y.~J.}\ \bibnamefont {Uemura}}, \bibinfo {author}
  {\bibfnamefont {R.~M.}\ \bibnamefont {Fernandes}}, \ and\ \bibinfo {author}
  {\bibfnamefont {N.}~\bibnamefont {Ni}},\ }\href
  {http://arxiv.org/abs/1603.04899} {\  (\bibinfo {year}
  {2016}{\natexlab{b}})},\ \Eprint {http://arxiv.org/abs/1603.04899}
  {arXiv:1603.04899} \BibitemShut {NoStop}%
\end{thebibliography}

\end{document}